# ON THE CONNECTION BETWEEN GAMMA AND RADIO RADIATION SPECTRA IN PULSARS

V. M. Kontorovich, A. B. Flanchik

Institute of Radio Astronomy of NAS of Ukraine, Kharkov 61002, Chervonopraporna St 4

e-mail: vkont@ira.kharkov.ua, alex_svs_fl@vk.kh.ua

The model of pulsar radio emission is discussed in which a coherent radio emission is excited in a vacuum gap above polar cap of neutron star. Pulsar X and gamma radiation are considered as the result of low-frequency radio emission inverse Compton scattering on ultra relativistic electrons accelerated in the gap. The influence of the pulsar magnetic field on Compton scattering is taken into account. The relation of radio and gamma radiation spectra has been found in the framework of the model.

PACS: 97.60.Gb, 41.20.Jb

## 1. INTRODUCTION

Mechanisms responsible for low-frequency radiation in pulsars have long been known [1]. However, they include a difficulty in explaining coherence [2] necessary for the appearance of high brightness temperatures $T_{BR} \sim 10^{25} - 10^{28} K$, typical for the radio radiation from pulsars [3].

A mechanism of coherent radio radiation, discussed by us, consists in exciting of oscillations in a "cavity resonator" [4] (Fig. 1), which is a vacuum gap. These oscillations are excited by electron bunches flowing from needles of the surface of the pulsar polar cap in the discharge process [5]. They can be microscopic bulges on the solid surface or wave peaks on the liquid surface of the polar cap in a high longitudinal electric field appearing in the gap [6].

In traditional mechanisms of pulsar radio radiation, bunches necessary for coherent radiation cannot be formed, because relativistic electron beams rapidly pass through oscillation generation regions in the inhomogeneous pulsar magnetosphere plasma. Conditions in discharges in the gap can be different. Since discharges occur "randomly" at different places of the polar cap, time is sufficient to form a next bunch: it is not limited by small times of acceleration and passage of an electron through the gap.

In this work, we consider the connection of radio radiation appearing in the gap with X- and gamma rays from a pulsar that are generated due to the inverse Compton scattering[1] of low-frequency oscillations on ultra relativistic electrons in the gap. As will be shown the hard radiation carries important information about oscillations in the gap. Note that the inverse Compton scattering was applied many times to explain radiation from pulsars, but it was considered on thermal photons of radiation from the polar cap [7] or on synchrophotons near the light cylinder [8]. The "after" Compton scattering with small frequency change was used to explain coherent radio radiation in the "porous" pulsar magnetosphere [9].

---

[1] In this version we take into account the magnetic field influence on the Compton scattering, which was not been taken into account in [30].

We consider here only the radiation emitted from the cavity through a "waveguide" appearing near the magnetic axis of the pulsar [10] (and we hope that received estimations will be valid for the more general case of radiation). The possibility of such radiation is corroborated for the three-humped shape of a pulse (as, e.g., from the PSR 0329-54 and 2045-15 pulsars), where the central hump is associated with radiation along the axis [11], whereas the other two humps are attributed to the maximum-curvature cone[2] [1]. Such a central part is also present in pulses from certain gamma pulsars [12]. Radio radiation emitted through other channels, including that penetrating through the magnetosphere plasma, can have an intensity on the order of the intensity of the radiation emitted through the waveguide (at least for some pulsars [9]). Note that the effective radii of the waveguide can be different for radio radiation and hard radiation due to its complex shape and spectrum.

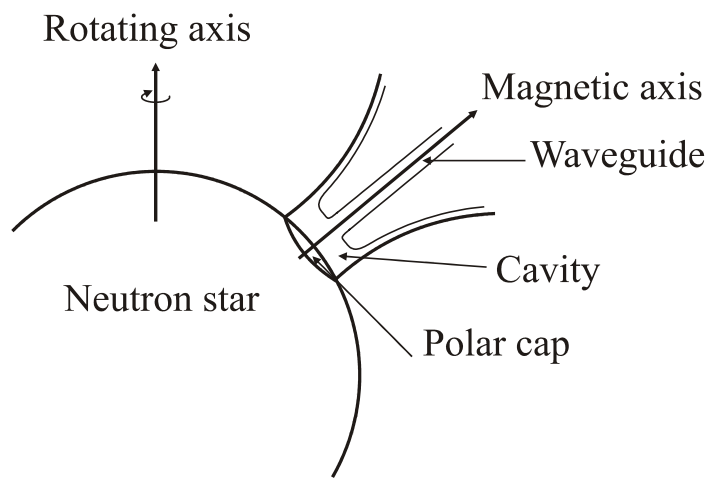

*Fig.1. "Resonator" and "waveguide" on the pulsar scheme in the open force line region.*

2. RADIO EMISSION IN A RESONATOR AND A WAVEGUIDE

Let us consider low-frequency oscillations excited by discharges in the vacuum gap of a pulsar.

---

[2] The formation of the cone and corresponding two-humped pulse shape is connected not only with the maximum-curvature field lines at the periphery of the polar cap, but also with geometric conditions of pair production, for which the angle of intersection of a photon with the field line in which a pair is produced must be sufficiently large [3]. Only photons emitted from the peripheral field lines can ensure such angles at short mean free paths (for a dipole magnetic field). For this reason, even in the absence of curvature photons, e.g., when Compton photons prevail, as in the case considered below, the complete cone, as well as the corresponding two-humped curve, can be formed.

The energy density $U$ of these oscillations can be estimated using the power $I_R$ of the radio radiation through the waveguide

$$I_R = c\, U_R\, \Sigma_w, \quad \Sigma_w \sim \pi R_w^2, \tag{1}$$

where $R_w$ is the radius of the waveguide for which various models exist (see, e. g., [10]). The sense of the index R in the energy density is to be shown bellow. The radius $R_w$ will be considered bellow as a parameter such that $R_w \ll R_{PC}$, where $R_{PC} \approx R_* \sqrt{\Omega R_* / c}$ is the polar cap radius, $R_* \sim 10^6\, cm$ is the neutron star radius, $\Omega = 2\pi/P$ is the angular velocity of pulsar rotation.

According to [1], the power of radio radiation from most pulsars is $I_R \sim 10^{27} - 10^{30}\, erg/s$, so that the energy density $U_R$ at the parameters $P = 0.1 s$, $R_w \sim 10^3\, cm$ is equal to

$$U_R \sim 10^{10} - 10^{13}\, erg/cm^3. \tag{2}$$

This energy density is much higher than the energy density $U_T \sim 10^6\, erg/cm^3$ of the thermal radiation from the pulsar polar cap (for the polar cap temperature $T_{PC} \sim 10^5\, K$).

The frequency of the radio radiation excited in the gap is bounded from below by the time-of-flight $\tau$ of accelerated electrons through the vacuum gap:

$$\omega \geq \frac{2\pi}{\tau} \approx \frac{2\pi c}{h}, \tag{3}$$

where $h$ is the gap height. Condition (3) for $h \sim 10^4\, cm$ provides $\omega \geq 10^6\, s^{-1}$.

Another low-frequency bound is associated with the condition of the yield of radio radiation through the waveguide $\lambda \leq R_w$ ($\lambda$ is the wavelength), or [3]

$$\omega \geq \frac{2\pi c}{R_w}. \tag{4}$$

The low-frequency cut-off of the radio spectrum $\omega_1$ is defined by the more hard condition from (3) and (4). For the parameters $P = 0.1 s$, $R_w \approx 10^3\, cm$ the bound (4) yields $\omega \geq 10^8\, s^{-1}$. It is important that the waveguide radius $R_w$ and gap height are functions of the pulsar parameters and can be strongly different. For fast rotating pulsars with $h < R_{PC}$ [3], the relation $R_w \sim h$ is possible and conditions (3) and (4) coincide with each other. In this case, even the longest wavelength modes can be emitted through the waveguide. When $R_w < h$, there are "locked" modes within the frequency range $\omega_{min} < \omega < \omega_1$ which are not transmitted through the waveguide. In this case, the energy density of the oscillations in the resonator can be much higher than value (2) based on the radio radiation (2). Therefore, estimate (2) is valid only when the contribution from the locked modes with $\lambda < R_w$ to the energy density is insignificant. The existence of the locked modes must also affect the intensity of hard radiation appearing due to the inverse Compton scattering.

---
[3] In general, on has to consider the activation character of the dispersion law in the waveguide and the resonator, but it does not act on the estimations.

In this work, the high-frequency bound is estimated using high-frequency cutoff in the radio radiation spectrum

$$\omega \leq \omega_{max} \sim 10^9 - 10^{10} \, s^{-1} \, . \qquad (5)$$

Let us note that the frequency $\omega_{max}$ has the order of the plasma frequency $\omega_{pl}$ determined by the Goldreich-Julian charge density $\rho_{GJ} = -\vec{\Omega} \cdot \vec{B}/(2\pi c)$ above the polar cap as $\omega_{pl}^2 = 4\pi e \rho_{GJ}/m$, it corresponds to the upper surface of the considered resonator. The frequency range in the spectrum of hard radiation appearing due to the inverse Compton scattering of radio radiation by electrons in the gap is linked (see below) with the frequencies $\omega_{min}$ and $\omega_{max}$.

Let us estimate the contribution of the locked modes to the energy density of the low-frequency oscillations in the gap. The total energy density $U$ of the low-frequency oscillations can be written as $U = U_{tr} + U_R$, where $U_{tr}$ is the contribution of the locked modes. For isotropic distributions we can write

$$U_{tr} = \int_{\omega_{min}}^{\omega_1} U(\omega) d\omega, \; U_R = \int_{\omega_1}^{\omega_{max}} U(\omega) d\omega \, . \qquad (6)$$

Assuming that the spectral power distribution of the pulsar radio emission has the form

$$I(\omega) \sim \omega^{-\alpha_R} \, , \qquad (7)$$

where $\alpha_R$ is the radio spectral index, we consider the distribution as isotropic with the same spectral index $U(\omega) \propto \omega^{-\alpha_R}$ in the whole frequency range $\omega_{min} \leq \omega \leq \omega_{max}$. In the case of falling spectrum $\alpha_R > 1$ for quantity $\mu = U_{tr}/U_R$ we have

$$\mu \approx \left(\frac{\omega_1}{\omega_{min}}\right)^{\alpha_R - 1} \left[1 - \left(\frac{\omega_{min}}{\omega_1}\right)^{\alpha_R - 1}\right] . \qquad (8)$$

The possibility of significant energy concentration in the locked modes can be seen for $\omega_{min} \ll \omega_1$. If $h > R_w$ then we have $\omega_{min} \sim 2\pi c/h$, $\omega_1 \sim 2\pi c/R_w$ and (8) can be estimated as

$$\mu \approx \left(\frac{h}{R_w}\right)^{\alpha_R-1} \approx 10^{\alpha_R-1}\left[\left(\frac{h}{10^4 cm}\right)\cdot\left(\frac{10^3 cm}{R_w}\right)\right]^{\alpha_R-1}. \tag{9}$$

The estimation of the locked modes contribution in the energy density of low-frequency field is

$$U_{tr} \approx 10^{\alpha_R-1} U_R \left[\left(\frac{h}{10^4 cm}\right)\cdot\left(\frac{10^3 cm}{R_w}\right)\right]^{\alpha_R-1}. \tag{10}$$

The average energy density $U$ can also be estimated from the conservation law for the electromagnetic field energy in the cavity excited by the external discharge currents. Actually, the field energy conservation law yields $div\ \vec{S} = -\overline{\vec{j}_{ex}\vec{E}_\sim}$, where $\vec{S}$ is the Poynting vector, $\vec{j}_{ex}$ is the density of the external currents of the sparks and $\vec{E}_\sim$ is strength of the low-frequency electric field, the bar means the averaging over time. Integrating this equation on the resonator volume and using the Gauss' theorem, we obtain (omitting the bars in other terms)

$$\oint \vec{S} d\vec{\Sigma} = -\int \overline{\vec{j}_{ex}\vec{E}_\sim}\, dV. \tag{11}$$

Note that the power emitted in the radio band is $I_R = \oint \vec{S} d\vec{\Sigma} = c U_R \Sigma_w$, where $U_R = U/(1+\mu)$, $U \sim E_\sim^2/(4\pi)$, and estimating the integral in the right side as

$$\int \overline{\vec{j}_{ex}\vec{E}_\sim}\, dV \leq \eta \cdot c\rho_{GJ} \cdot E_\sim \Sigma_{PC} h,$$

we find

$$U \leq 4\pi(1+\mu)^2 (\Sigma_{PC}/\Sigma_w)^2 (\eta \cdot h \cdot \rho_{GJ})^2. \tag{12}$$

Here $\eta < 1$ is the ratio of averaged density of spark current to the Goldreich-Julian current density $j_{GJ} = c\rho_{GJ}$. It follows from here under the parameters given above that $U \leq 10^{16} \eta^2 (1+\mu)^2\ erg/cm^3$.

Let us estimate the cavity Q-factor $Q = \omega/2\delta$, where $\delta$ is the damping decrement of the low-frequency field. This decrement can be expressed in terms of the energy loss due to radiation as $\delta = -\dot{W}/(2W) = c\Sigma_w/(2(1+\mu)\Sigma_{PC} h)$, where $W = U \cdot \Sigma_{PC} \cdot h$ is the electromagnetic oscillation energy in the cavity, $\dot{W} = -I_R = -cU\Sigma_w/(1+\mu)$ is the energy loss due to radiation through the waveguide, $\Sigma_{PC} = \pi R_{PC}^2$ is the polar cap area. Finally we have for the Q-factor

$$Q = (1+\mu)\frac{\omega}{c}\cdot h\cdot\frac{\Sigma_{PC}}{\Sigma_w} \approx 70(1+\mu)\left(\frac{\omega}{10^6 s^{-1}}\right)\cdot\left(\frac{h}{10^4 cm}\right)\cdot\left(\frac{10^3 cm}{R_w}\right)^2\left(\frac{1s}{P}\right) \tag{13}$$

For parameters $P = 0.1 s$, $R_w = 10^3 cm$, $h = 3 \cdot 10^4 cm$ and spectral index of the radio emission $\alpha_R = 3$ we have $\mu \sim 10^3$ and the Q-factor according to (13) can vary from $10^2$ to $10^6$.

### 3. THE COMPTON GAMMA RADIATION AND ITS POWER ESTIMATION

The inverse Compton scattering (Fig. 2) of the low-frequency radiation by ultrarelativistic electrons in the gap must cause powerful X and gamma radiation (Compton radiation). Let us determinate the scattered photon energy and estimate the power of this radiation.

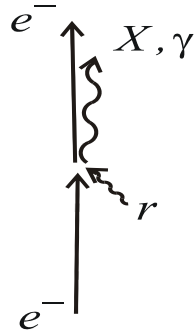

*Fig.2. Scheme of inverse Compton scattering. Here $r$ denotes a photon of radio emission excited in the gap, X or $\gamma$ denote X or gamma photon.*

The scattered photon energy $\hbar\omega_\gamma$ is given by

$$\hbar\omega_\gamma = \hbar\omega \frac{1 - \frac{V}{c}\cos\theta}{1 - \frac{V}{c}\cos\theta'}, \qquad (14)$$

where $V = c\sqrt{1 - \Gamma^{-2}}$ is the electron velocity, $\omega$ − initial radio emission frequency, $\theta, \theta'$ − angles between the momenta of initial and final photons and the magnetic field. Taking into account that because of relativistic aberration $1 - (V/c)\cos\theta \sim 1/\Gamma^2$, we obtain the frequency of scattered photon

$$\omega_\gamma = \omega \Gamma^2 (1 - \frac{V}{c}\cos\theta) \approx \omega \Gamma^2. \qquad (15)$$

This radiation belongs to X and gamma ranges.

As it is known [13], a strong magnetic field implies significantly on Compton scattering due to resonance and quantum effects. But for our frequency range and the field values such effects are as a rule negligible. However, as it was shown by Blandford and Scharlemann [14], a Compton scattering on ultra relativistic electrons oscillating strictly along the very high magnetic field is strongly suppressed due to transverse character of waves and relativistic aberration. More important (but also suppressed) is the Compton scattering due to electron drift oscillations [14] in the pulsar magnetic field and electric field of low-frequency radiation (Fig. 3).

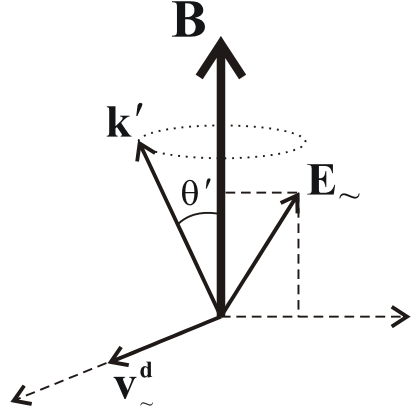

*Fig. 3. The scheme of electron drift oscillation in the pulsar magnetic field $\vec{B}$ and electric field $\vec{E}_\sim$ of low-frequency radiation in the gap, $\vec{V}_\sim^d = c[\vec{E}_\sim, \vec{B}]/B^2$ is the oscillation drift velocity.*

The differential cross-section of Compton scattering in the electron rest frame with the drift influence (see Fig. 3) is given by [14]

$$d\sigma = \frac{r_e^2}{4}\frac{\omega^2}{\omega_B^2}(1+\cos^2\theta)(1+\cos^2\theta')d\Omega', \quad d\Omega' = 2\pi\sin\theta'd\theta', \qquad (16)$$

where $\omega_B = eB/(mc)$, $r_e = e^2/(mc^2)$. Using the Lorentz transformations for angles and frequency, we obtain the scattering cross-section on ultrarelativistic electron in the laboratory frame

$$d\sigma = r_e^2\frac{\omega^2}{\omega_B^2}\frac{(1-\frac{V}{c}\cos\theta)^2}{(1-\frac{V}{c}\cos\theta')^2}d\Omega'. \qquad (17)$$

The denominator in (17) corresponds to relativistic aberration. In the region of small angles $\theta' \sim 1/\Gamma$, giving the main contribution into scattering, we have $1-\frac{V}{c}\cos\theta' \sim 1/\Gamma^2$, so in this region $d\sigma/d\Omega' \sim \Gamma^4$.

Integrating (17) over $d\Omega'$, we find the total scattering cross-section

$$\sigma = \sigma_T\frac{\omega^2\Gamma^2}{\omega_B^2}(1-\tfrac{v}{c}\cos\theta)^2 \quad \text{(L-frame)}, \qquad (18)$$

that can be written in relativistic-invariant form

$$\sigma = \sigma_T\frac{(kp)^2}{m^2\omega_B^2}, \qquad (19)$$

where $p = \left(\frac{\varepsilon}{c}, \vec{p}\right)$, $k = \left(\frac{\omega}{c}, \vec{k}\right)$ are 4-momentum of electron and wave 4-vector of initial photon and their scalar product is $pk = m\Gamma\omega(1-\frac{V}{c}\cos\theta)$. In the electron rest frame we have $p = (mc, 0)$, $pk = m\omega$ and the total scattering cross-section is given by

$$\sigma = \sigma_T \frac{\omega^2}{\omega_B^2} \qquad \text{(R-frame)}, \qquad (20)$$

it is the same that is obtained from (16) after integrating over the angles.

The probability $w(\vec{k}',\vec{k})d^3k'$ of Compton scattering of radio emission photon with emitting of gamma photon with wave vector in the interval $d^3k'$ can be written as [15]

$$w(\vec{k}',\vec{k})d^3k' = c\frac{d\sigma}{d\Omega'}\left(1-\frac{V}{c}\cos\theta\right)d\Omega',$$

Using (17), we obtain

$$w(\vec{k}',\vec{k}) = \frac{c^4 r_e^2}{\omega'^2} \frac{\omega^2}{\omega_B^2} \frac{(1-\frac{V}{c}\cos\theta)^3}{(1-\frac{V}{c}\cos\theta')^2} \delta\left(\omega' - \omega\frac{1-\frac{V}{c}\cos\theta}{1-\frac{V}{c}\cos\theta'}\right). \qquad (21)$$

Let us estimate the total power $I_\gamma$ of the Compton radiation

$$I_\gamma = \int\int q(\Gamma)f(\Gamma,\vec{r})d\Gamma\, d^3x, \qquad (22)$$

where $f$ is the distribution function of electrons accelerated by the longitudinal electric field in the gap, $q$ is the power of the Compton radiation per particle. Taking into account the influence of the pulsar magnetic field and low-frequency oscillations excited in the gap, the electron energy losses in the resonator due to the scattering on low-frequency modes with the power-law spectrum (6) is given by [14]

$$q(\Gamma) = \int \hbar\omega'\, w(\vec{k}',\vec{k}) N_R(\vec{k}) \frac{2d^3k}{(2\pi)^3} d^3k' = cg\sigma_T U\Gamma^4, \qquad (23)$$

where in our case

$$g = \frac{24}{5} \cdot \frac{\alpha_R - 1}{\omega_B^2} \omega_{min}^{\alpha_R-1} \int_{\omega_{min}}^{\omega_{max}} \omega^{2-\alpha_R} d\omega, \quad \omega_B = \frac{eB}{mc}, \qquad (24)$$

and $N_R(\vec{k})$ implies the distribution function of the radio emission photons that can be written as $N_R(\vec{k}) = const \cdot \omega^{-(3+\alpha_R)}$. Normalizing it to the wave energy density in the resonator

$$U = \int \hbar\omega N_R(\vec{k}) \frac{2d^3k}{(2\pi)^3}$$

we obtain

$$N_R(\vec{k}) = \frac{\pi^2 c^3}{\hbar}(\alpha_R - 1) U \omega_{min}^{\alpha_R-1} \cdot \omega^{-(3+\alpha_R)}. \qquad (25)$$

Let us note that in the absence of the influence of the magnetic field (and the electron drift) on Compton scattering the cross-section would be given by the Thomson formula and the Compton energy losses would be [16] $q(\Gamma) = \frac{4}{3}c\sigma_T U\Gamma^2$ instead of (23). Comparing with (23), we see that in the presence of the strong magnetic field the Thomson cross-section must be replaced accordingly

$$\sigma_T \to \frac{3}{4}g\sigma_T\Gamma^2, \quad g \sim 2\cdot 10^{-14}\cdot 10^{-4\alpha_R}\cdot \left(\frac{\omega_{min}}{10^6 s^{-1}}\right)^{\alpha_R-1}\left(\frac{\omega_{max}}{10^{10} s^{-1}}\right)^{3-\alpha_R}\left(\frac{10^{12}G}{B}\right)^2. \quad (26)$$

The small parameter $(\omega/\omega_B)^2$ in the cross-section of Compton scattering (16) makes the Compton energy losses (23) much smaller than in the case of scattering on free electrons. This smallness in scattering may be compensated by both the factor $\Gamma^2$ and higher value of the energy density $U$ of the low-frequency oscillations in the gap, compatible with the processes of the particle acceleration up to Lorentz factor values $\Gamma \approx 10^7 - 10^8$.

Indeed, the distribution function $f$ can be written as

$$f(\Gamma,\vec{r}) = n_e\,\delta(\Gamma - \Gamma(z)), \quad (27)$$

($n_e$ is the electron concentration in the gap), where the electron Lorentz factor $\Gamma(z)$ obeys the equation of motion

$$\frac{d\Gamma(z)}{dz} = \frac{e}{mc^2}E_{\|}(z) - \frac{2e^2}{3mc^2 R_c^2}\Gamma^4(z) - g\frac{\sigma_T U}{mc^2}\Gamma^4(z), \quad (28)$$

where $E_{\|}(z)$ is the longitudinal electric field in the gap for which there are various models, $R_c = 7\cdot 10^7 cm \cdot \sqrt{P\,\sec}$ is the curvature radius of magnetic force line. In Eq. (28) the first term describes the electron acceleration in the field $E_{\|}(z)$, the second and third ones describe the energy losses due to curvature radiation and inverse Compton scattering. Substituting (23) and (27) into (22), we find

$$I_\gamma \approx cg\sigma_T n_e \cdot U\Sigma_w\int \Gamma^4(z)dz \sim cg\sigma_T n_e\cdot U\Sigma_w h\cdot\bar{\Gamma}^4 \quad (29)$$

for the power of the hard radiation exiting the waveguide, where $\bar{\Gamma}$ is the maximum value of electron Lorentz factor achieved in the gap and determined by the most effective mechanism of energy loss. When the condition

$$U \gg U_{min} = \frac{2e^2}{3R_c^2 g\sigma_T} = 2.3\cdot 10^{14}\left(\frac{10^8 cm}{R_c}\right)^2\left(\frac{10^{-25}}{g}\right)\frac{erg}{cm^3} \quad (30)$$

takes place, the energy losses due to inverse Compton scattering predominate over the curvature radiation losses. In this case one can omit the second term in the right side of Eq. (28) and the maximum Lorentz factor in the gap is determined by inverse Compton scattering. Its estimation is given by

$$\overline{\Gamma} \approx \left( \frac{mc^2}{g\sigma_T U z_\Gamma} \right)^{\frac{1}{3}}, \qquad (31)$$

where $z_\Gamma$ is the height above the pulsar surface at which the maximum of $\Gamma$ is achieved. The height $z_\Gamma$ can be obtained from the equation

$$eE_{\parallel}(z_\Gamma) \cdot z_\Gamma^{4/3} = \left( \frac{(mc^2)^4}{g\sigma_T U} \right)^{\frac{1}{3}}. \qquad (32)$$

As a result the ratio of the power of the pulsar Compton radiation to the power of its radio emission can be estimated using (1), (8) and (23) as

$$\frac{I_\gamma}{I_R} \approx \mu g \cdot \sigma_T n_e h \cdot \overline{\Gamma}^4. \qquad (33)$$

|  | B0531+21 Crab | B0833-45 Vela | B1951+32 |
|---|---|---|---|
| $I_R$, erg/s | $4 \cdot 10^{31}$ | $1.6 \cdot 10^{29}$ | $8 \cdot 10^{28}$ |
| $I_\gamma$, erg/s | $5 \cdot 10^{35}$ | $8.6 \cdot 10^{33}$ | $2.5 \cdot 10^{34}$ |
| $I_\gamma / I_R$ | $1.25 \cdot 10^4$ | $5.37 \cdot 10^4$ | $3.15 \cdot 10^5$ |

Tab.1. Observed powers of three pulsars in the radio [1] and gamma range [12].

The observed powers of radio emission and gamma radiation for three pulsars are given in Tab.1, from which one can see that $I_\gamma / I_R \sim 10^4 - 10^5$. In our model such value of ratio $I_\gamma / I_R$ can be obtained only due to the locked modes contribution with the $\mu \approx 10^3$. Such value of $\mu$ can be obtained for parameters $h = 2 \cdot 10^4 cm$, $\overline{\Gamma} = 3 \cdot 10^8$, $n_e = 5 \cdot 10^{12} cm^{-3}$, $g = 2 \cdot 10^{-26}$.

The minimum $\omega_{\gamma 1}$ and maximum $\omega_{\gamma 2}$ frequencies of the Compton gamma radiation can be estimated by the frequency interval of radio emission in the gap

$$\omega_{\gamma 1} \approx \omega_{min} \overline{\Gamma}^2, \quad \omega_{\gamma 2} \approx \omega_{max} \overline{\Gamma}^2. \qquad (34)$$

The estimations of electron Lorentz factor $\overline{\Gamma}$ and frequency interval of the Compton radiation depend on the model of the longitudinal electric field in the gap. In the model with suppressed exit of particles from the star surface [6] the longitudinal field takes a form $E_{\parallel}(z) = \Omega B(h-z)/c$, where $B$ is the magnetic field near the surface of star. The maximal electron Lorentz factor, the acceleration length and Compton gamma radiation frequencies can be estimated in this case using (31), (32) and (34) as

$$\overline{\Gamma} \approx 3.0 \cdot 10^7 \left[ \left( \frac{10^{-25}}{g} \right) \cdot \left( \frac{10^{16} erg/cm^3}{U} \right) \cdot F \right]^{\frac{1}{6}}, \quad (35)$$

$$z_{\Gamma} \approx 4.8 \cdot 10^4 cm \left[ \frac{1}{F} \cdot \left( \frac{10^{-25}}{g} \right) \cdot \left( \frac{10^{16} erg/cm^3}{U} \right) \right]^{\frac{1}{2}}, \quad (36)$$

$$\omega_{\gamma} \approx 10^{15} \omega \cdot \left[ \left( \frac{10^{-25}}{g} \right) \cdot \left( \frac{10^{16} erg/cm^3}{U} \right) \cdot F \right]^{\frac{1}{3}}, \quad (37)$$

where

$$F = \left( \frac{1s}{P} \right) \cdot \left( \frac{R_c}{10^7 cm} \right) \cdot \left( \frac{10^{10} s^{-1}}{\omega_{max}} \right).$$

The longitudinal electric field in the model with free exit [17] (with general relativity corrections [3]) is given by

$$E_{\parallel}(z) = \frac{3 \Omega B}{4 c R_*} a z (h - z),$$

where $a = 4(R_g/R_*)^3 \cos\chi + \sqrt{\Omega R_*/c} \sin\chi \cos\varphi_m$, $R_g \sim 0.2 R_*$ is the neutron star gravitational radius, $\chi$ is an angle between the magnetic axis and the rotation one, $\varphi_m$ is an azimuth angle relatively to the pulsar magnetic axis. Analogously to (35) − (37) we find

$$\overline{\Gamma} \approx 2.1 \cdot 10^7 \left[ \left( \frac{10^{-25}}{g} \right)^2 \left( \frac{10^{16} erg/cm^3}{U} \right)^2 \cdot a F \right]^{\frac{1}{9}}, \quad (38)$$

$$z_{\Gamma} = 1.5 \cdot 10^5 cm \left[ \frac{1}{a F} \cdot \left( \frac{10^{-25}}{g} \right) \cdot \left( \frac{10^{16} erg/cm^3}{U} \right) \right]^{\frac{1}{3}}, \quad (39)$$

$$\omega_{\gamma} \approx 4.4 \cdot 10^{14} \omega \left[ \left( \frac{10^{-25}}{g} \right)^2 \left( \frac{10^{16} erg/cm^3}{U} \right)^2 \cdot a F \right]^{\frac{2}{9}} \quad (40)$$

For low-frequency emission satisfying the conditions (3) and (5) the Compton radiation belongs to X and gamma regions of the spectrum.

It is necessary to note that in the considered case the effective acceleration of electrons is possible under much higher energy density of low-frequency emission in the gap than under scattering on free electrons without the influence of magnetic field. Due to this and because of the locked modes contribution the energy density of low-frequency oscillations in the gap can be significantly higher than its estimation (2) based on the pulsar radio emission.

## 4. SPECTRAL DISTRIBUTION OF COMPTON GAMMA RADIATION

Let us consider the spectral distribution of the gamma radiation resulting from inverse Compton scattering of the radio emission by electrons in the gap. It can be expressed as

$$I_\gamma(\omega, z) = \frac{2\hbar\omega^3}{(2\pi c)^3} \int_{\Sigma_w} dx\, dy \int d\Omega'\, N_\gamma(\vec{k}, \vec{r}) \tag{41}$$

where $N_\gamma(\vec{k}, \vec{r})$ is extremely anisotropic distribution function of the Compton photons which is given by the kinetic equation

$$c\frac{\partial}{\partial z} N_\gamma(\vec{k}_\gamma, \vec{r}) = \int d\Gamma\, d^3k \cdot w(\vec{k}_\gamma, \vec{k}) f(\Gamma, \vec{r}) N_R(\vec{k}, \vec{r}). \tag{42}$$

From (41) and (42) we find

$$I_\gamma(\omega_\gamma, z) = \frac{2\hbar\omega_\gamma^3}{(2\pi c)^3 c} \int_{\Sigma_w} dx\, dy \int d\Omega' \int_0^z dz' \int d\Gamma\, d^3k \cdot w(\vec{k}_\gamma, \vec{k}) f(\Gamma, x, y, z') N_R(\vec{k}, x, y, z'). \tag{43}$$

Integrating (42) we have supposed that $N_\gamma\big|_{z=0} = 0$, neglecting the thermal X-photons [18] emitted by the hot polar cap of the star. Substituting (21), (25) and (27) into (43) we obtain

$$I_\gamma(\omega, z) \approx c\sigma_T \frac{\omega^{2-\alpha_R}}{\omega_B^2}(\alpha_R - 1)\omega_{min}^{\alpha_R - 1} n_e U \Sigma_w \int_0^z \Gamma^{2\alpha_R - 2}(z')\, dz' \tag{44}$$

for the spectrum of the gamma radiation. From (44) one can see the connection between indexes of spectral distributions of low-frequency (radio) and Compton (gamma) photons:

$$\alpha_\gamma = \alpha_R - 2, \quad I_\gamma(\omega) \sim \omega^{-\alpha_\gamma}, \quad I_R(\omega) \sim \omega^{-\alpha_R}. \tag{45}$$

The comparing of the EGRET data [19] on pulsar gamma radiation and the data on their radio emission is given in Tab. 2-4. For PSR B0531+21 (Crab), B0833-45 (Vela) and J0633-1746 (Geminga) one sees a satisfactory coincidence with obtained connection of spectral indexes (45).

| PSR | $\alpha_R$ | $\alpha_\gamma$ | $\alpha_R - \alpha_\gamma$ |
|:---:|:---:|:---:|:---:|
| Crab B0531+21 | 2.9 ± 0.4 [20] | 1.07 ± 0.03 [19] | 1.83 ± 0.4 |
| Vela B0833-45 | $2.9^{+1.1}_{-0.8}$ [21] | 0.54 ± 0.01 [19] | $2.36^{+1.1}_{-0.8}$ |
| Geminga J0633-1746 | 2.65 ± 0.2 [22] | 0.39 ± 0.02 [19] | 2.26 ± 0.2 |

Tab.2. Pulsar spectral indexes in radio and gamma emissions with index difference about 2.

Spectral indexes of other known pulsars are given in Tab. 3 and 4. It is seen that for the two pulsars from Tab. 3 the difference $\alpha_R - \alpha_\gamma$ approaches 1 and for the three pulsars from Tab. 4 the spectral indexes have close values $\alpha_R \sim \alpha_\gamma$, which does not correspond the applied model of emission (scattering) [4].

| PSR | $\alpha_R$ | $\alpha_\gamma$ | $\alpha_R - \alpha_\gamma$ |
|---|---|---|---|
| J0218+4232 | 2.92 [24] | 1.6 [23] | 1.32 |
| B1951+32 | 1.6 [24] | 0.78 ± 0.09 [19] | 0.82 ± 0.09 |

Tab.3. Pulsar spectral indexes in radio and gamma emissions with the index difference near 1.

| PSR | $\alpha_R$ | $\alpha_\gamma$ | $\alpha_R - \alpha_\gamma$ |
|---|---|---|---|
| B1509-58 | 0.37 [25] | 0.68 ± 0.09 [19] | -0.31 ± 0.09 |
| B1046-58 | 0.96 [26] | 1.0 ± 0.10 [27] | -0.04 ± 0.10 |
| B1706-44 | 0.53 [23] | 0.56 ± 0.05 [19] | -0.03 ± 0.05 |

Tab.4. Pulsar spectral indexes with close index values in radio and gamma ranges

The pulsar B1706-44 (Tab. 4) in the gamma spectrum may have a kink under photon energy $\hbar\omega = 1 GeV$ [28] then its spectral index increases by 1: $\alpha_\gamma = 0.27 \pm 0.09$ for $\hbar\omega < 1\,GeV$ and $\alpha_\gamma = 1.25 \pm 0.13$ for $\hbar\omega > 1\,GeV$. In the model under consideration the kink can arise due to the transition from the scattering on the locked modes to the scattering on the radio emission when $\hbar\omega_1 \overline{\Gamma}^2 \sim 1 GeV$, where $\omega_1 \approx 2\pi c / R_w$. For the Lorentz factor $\overline{\Gamma} \sim 10^8$ such transition appears at the frequency $\omega_1 \sim 10^8 s^{-1}$ which corresponds to the waveguide radius $R_w \sim 10^3 cm$.

For the pulsar B1055-52 there are no data on the radio emission spectrum, the corresponding spectral index in gamma-rays is $\alpha_\gamma = 0.50 \pm 0.13$ [19]. For considered emission mechanism the spectral index in radio according to (45) may be equal $\alpha_R = \alpha_\gamma + 2 \approx 2.5 \pm 0.13$.

## 5. CONCLUSIONS

In this work we suggest a model of pulsar emission in which a powerful coherent emission arises in a vacuum gap above the star polar cap. The gap is considered as a resonator [4], excited by sparks in a longitudinal electric field. The pair production is suppressed near the magnetic axis because of low curvature of magnetic force lines that leads to the formation of waveguide through which the radio emission leaves the resonator[5]. The spectral properties of the radio emission can be related with resonator and waveguide parameters. Inverse Compton scattering of the radio emission on the electrons in the gap on the one hand leads to X and gamma radiations of pulsars through the waveguide, and on the other hand it restricts the achieved energies of particles accelerated in the gap. That is why the gamma radiation and the radio emission must be linked each other. Some observational consequences of this relation have been predicted. In particular, the minimum and maximum frequencies of the Compton radiation according to (37) or (40) correlate with the frequency interval of the radio emission and must depend on the energy density of the low-frequency emission in the resonator. The relation (45) between the spectral indexes of power radiation in radio and gamma frequency ranges has been obtained. The ratio of the gamma to radio power radiation (33) and the relation of spectral indexes (45) do not defy the observation data for a number of pulsars.

---

[4] The case $\alpha_R \sim \alpha_\gamma$ might arise when electromagnetic oscillations are excited in the pulsar outer gap [29, 1], where the magnetic field is much weaker than near the star surface. In this case $\omega > \omega_B$ and the scattering cross-section $\sigma \sim \sigma_T$ would not depend on the frequency that leads to $\alpha_\gamma = \alpha_R$. The case $\alpha_R - \alpha_\gamma = 1$ might correspond to intermediate situation $\omega \sim \omega_B$ in the laboratory frame.

The locked modes play an important role in the considered model. That is the part of low-frequency radiation in the gap unable to pass the waveguide if the condition (4) does not take place. Their contribution to the energy density of low-frequency emission is determined by the ratio of the gap height to the waveguide radius. One can see from (8) and (9) a considerable concentration of energy in the locked modes is possible that may affect on the power of the Compton gamma radiation and its connection with the pulsar radio emission. In general, the spectral index for the locked modes may differ from its value $\alpha_R$ for the observed radio emission. In this case a kink in the pulsar gamma radiation spectrum may appear.

We thank V. S. Beskin, I. F. Malov, S. A. Petrova and O. M. Ul'yanov for the important discussions and notices. We also thank S. A. Petrova for pointing to R. Blandford and E. Scharlemann paper [14].

REFERENCES

1. I. F. Malov, Radiopulsars, Moscow, "Nauka", 2004, 192 p. (in Russian).

2. V. L. Ginzburg, Uspekhi Fiz. Nauk, **103**, 393 (1971); V. L. Ginzburg, V. V. Zheleznyakov, V. V. Zaitsev, Uspekhi Fiz. Nauk, **98**, 201 (1969).

3. V. S. Beskin, Uspekhi Fiz. Nauk, 169, 1169 (1999); *Axial symmetry stationary in astrophysics*, Moscow, "Fizmatlit" 2006, 384 p. (in Russian).

4. V. M. Kontorovich, Radiophysics and Radioastronomy, **11**, 308 (2006).

5. V. S. Beskin, Astron. Zh., **59**, 726 (1982).

6. M. A. Ruderman, P. G. Sutherland, Astrophys. J, **196**, 51 (1975).

7. N. S. Kardashev, I. G. Mitrofanov, I. D. Novikov, Astron. Zh., **61**, 1113 (1984).

---

[5] There are also the possibilities of the radio emission passing through the magnetosphere plasma and its leaving through the slot gap which require a special consideration.

8. I. F. Malov, G. Z. Machabeli, Astron. Zh., **79**, 755 (2002).

9. G. J. Qiao, W. P. Lin, Astron. & Astrophys, **333**, 172 (1998).

10. V. S. Beskin, A. V. Gurevich, Ya. N. Istomin, Zh. Eksp. Theor. Fiz., **150**, 257 (1986).

11. J. M. Rankin, Astrophys. J, **274**, 333 (1983).

12. D. J. Thompson, ArXiv: astro-ph/0312272 (2003).

13. X. Y. Xia, G. J. Qiao, X. J. Wu, Y. Q. Hou, Astron. Astrophys., **152**, 93 (1985).

14. R. D. Blandford, E.T. Scharlemann, MNRAS, **174**, 59 (1976).

15. V. B. Berestetskii, E. M. Lifshitz, L. P. Pitaevskii, *Quantum electrodynamics,* Moscow, "Nauka" , 1979, 703 p. (in Russian).

16. Yu. P. Ochelkov, O. F. Prilutskii, I. L. Rosental, V. V. Usov, *Relativistic hydrodynamics and kinetics*, Moscow, "Atomizdat", 1979, 200 p. (in Russian).

17. J. Arons, E. T. Scharlemann, Astrophys. J, **231**, 854 (1979).

18. V. M. Kaspi, M. Roberts, A. K. Harding, astro-ph/0402136 (2004).

19. P.L. Nolan et al., Astron. Astrophys. Suppl. Ser. **120**, 61 (1996).

20. J. M. Rankin, J. M. Comella, H. D. Craft et al., Astrophys. J, **162**, 707 (1970).

21. W. Sieber, Astron. & Astrophys. **28**, 237 (1973).

22. M. A. McLaughlin, J. M. Cordes, T. H. Hankins, D. A. Moffett, ArXiv:astro-ph/9912410 (1999).

23. L. Kuiper, W. Hermsen, F. Verbunt et al., Astron. Astrophys., **359**, 615 (2000).

24. O. Maron, J. Kijak, M. Kramer et al., Astron. Astrophys. Suppl. Ser.,**147**, 195 (2000).

25. ATNF Pulsar Database http://www.atnf.csiro.au/research/pulsar/psrcat, Manchester, R. N., Hobbs, G. B., Teoh, A. & Hobbs, M., AJ, **129**, 1993-2006 (2005).

26. S. Johnston, A. G. Lyne, R. N. Manchester et al., MNRAS, **255**, 401 (1992).

27. V. M. Kaspi, J. R. Lackey, J. Mattox et al., Astrophys. J, 528, **445** (2000).

28. D. J. Thompson, M. Bailes, D. L. Bertsch et al., Astrophys. J., **465**, 385 (1996).

29. N. J. Holloway, Nature Phys. Sci., 246, 6 (1973).

30. V. M. Kontorovich, A. B. Flanchik, Pis'ma v Zh. Eksp. Teor. Fiz., **85**, 323 (2007).